# An enhanced Multipath Strategy in Mobile Ad hoc Routing Protocols

Zeyad M. Alfawaer
Department of computer science,
CCD,University of Dammam,
Dammam, Saudi Arabia
zmalfawaer@uod.edu.sa

Belgaum Mohammad Riyaz
Department of computer science,
CCS, AMA International University,
Salmabad, Bahrain
bmdriyaz@amaiu.edu.bh

*Abstract*— **The various routing protocols in Mobile Ad hoc Networks follow different strategies to send the information from one node to another. The nodes in the network are non static and they move randomly and are prone to link failure which makes always to find new routes to the destination. This research mainly focused on the study of the characteristics of multipath routing protocols in MANETS. Two of the multipath routing protocols were investigated and a comparative study along with simulation using NS2 was done between DSR and AODV to propose an enhanced approach to reach the destination maintaining the QoS. A possible optimization to the DSR and AODV routing protocols was proposed to make no node to be overburdened by distributing the load after finding the alternate multipath routes which were discovered in the Route discovery process. The simulation shows that the differences in the protocol highlighted major differences with the protocol performance. These differences have been analyzed with various network size, mobility, and network load. A new search table named Search of Next Node Enquiry Table (SONNET) was proposed to find the best neighbor node. Using SONNET the node selects the neighbor which can be reached in less number of hops and with less time delay and maintaining the QoS.**

## I. INTRODUCTION

In the present era of communication of wireless networks, research in MANETS has been progressing at a great speed, even is getting more popularity. The key reason for this increased attention is because most of the multimedia applications run in an infrastructure less environment. As the environments is infrastructure less and because of the nodes in this environment are not fixed, it becomes very difficult to provide a safe and secure environment in MANET. Being the nodes non static they follow random way mobility [1] which are connected by wireless links. MANETS are prone to frequent link failures and high mobility.

The topology being non static, the nodes are moving and being not stable because of the radio propagation range being limited. And more over the communication of nodes in MANET is done either through single hop or multihop transmissions. Each node in MANETS plays a dual role as hosts and routers. MANETS applications include communication at the borders in the battlefield, recovery of disaster, management of traffic by vehicular communication and to extend in the field of wireless networks. The nodes role is to know the optimal path to reach the destination and forward the information received to the sink node. But because the resources are very limited for the nodes and the wireless links being unreliable[2] it puts forward many challenges for designing efficient routing protocols. The routing algorithms that select an optimal path[3] during finding route to reach the destination make discovery of routes very frequently and thus this results with throughput being low and increased overhead.

The simulation shows that the differences in the protocol highlighted major differences with the protocol performance. These differences have been analyzed with various network size, mobility, and network load. A new search table named Search of Next Node Enquiry Table (SONNET) was proposed to find the best neighbor node. Using SONNET the node selects the neighbor which can be reached in less number of hops and with less time delay and maintaining the QoS.

## II. REVIEW OF RELATED LITERATURE AND STUDIES

In the early 1970's since the invent of DARPA packet radio networks [5], so many types of secure routing protocols have been designed for MANETS [6-11]. And the characteristics of the routing protocols are studied from [12].These protocols can be categorized as shown in fig. 1.

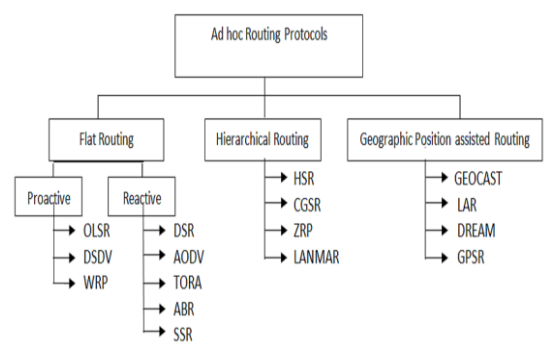

Fig. 1. Categorization of Ad hoc routing protocols.

*A. Reactive Routing Protocols*

1. DSR

The Dynamic Source Routing protocol [13] was explained as a reactive routing protocol. It has two phases: Discovery of Routes and Maintenance of Routes. In first phase of DSR, i.e discovery of routes, the route cache is verified for transmitting of packets to target, if there exists is no route to that target host, the initiation of Route Discovery is done by the source node. The broadcast message as a (Route Request)RR packet, adding the target and the initiator's identification is sent by the node. All the nodes that receive the RR packet in duplicate just discards the RR packet if it is sent by the same node. Otherwise, if it is a new request then the receiver node adds its identity to the list in the RR and rebroadcast it. Finally when RR packet reaches the target node, a Route Reply is delivered following the same route to the source which initiated the ROUTE REQUEST with a list of all addresses that were visited between source and destination. This established new route is added to Route Cache. In the second phase i.e Maintenance of Route the path through which a packet was sent to the destination is to be maintained if that route is broken as the topology being dynamic, the nodes move randomly far away.

After the discovery of route, considering the next node's address mentioned the source node(S) sends packet, and even gives the message after the packet has been received. This confirmation can be given by the node by using a passive acknowledgment, a link layer acknowledgment. If a node cannot make the confirmation after some finite number of retransmission of packet, then a error message in the form of ROUTE ERROR will be received by the source specifying that the link between the two successive nodes is broken. This broken link is erased from the Route Cache so that this path is not used for transmission of packets further.

2. AODV

The AODV protocol defined in [14], explains its functionality. To forward packet, if no path is found in the routing table, then as a broadcast a Route Request (RREQ) is sent. All the receivers who receive this Route Request will form the routes to the node(S) along the same path through which they received the Route Request message. Along the active route between S and T nodes, then the node(S) receive a Route Reply (RREP) as unicast. If not then, still the RREQ message will be re-broadcasted. In order to stop the broadcasting of the same request repeatedly, each request sent has a unique identification as NODE_ ID, Broadcast couple_ ID. Each node maintains a track of the various processed RREQs. The route re-discovery process is initiated when the nodes receive Route Error packets from their neighbor node when there are broken paths. A destination sequence count is added to prevent routing loop and find the fresh route. The sequence count of a mobile node can only be updated in the increasing mode by itself monotonically. If the sequence number is large, it is a fresh route. In both the Request message and Reply message, the sequence number is incremented. In order to stop the source node to use a stale path the sequence number in Route Reply must be computed.

It must be compared with the respective Route Request. The greatest destination sequence count is used when there is more than one path represented by different Route Replies. The shortest path is chosen if there is more than one path with equal sequence count. The most admirable features of AODV are; In the static networks it has low byte overhead and by using destination sequence count the routing is free of loop.

3. TORA

The Temporally ordered Routing Algorithm in[15] explains as follows. It is designed considering the link reversal concept and is an algorithm free of loops. This protocol works in are three phases as follows: (a) The first phase starts with creating a route (b) the second phase goes with maintaining the route and (c) the third phase removes all the invalid routes. During the functions of creating and maintaining the routes, the mobile hosts uses a "height" metric. TORA's metric has five elements, namely :(i) link failure's time, (ii)host's unique ID showing the level, (iii)a bit for indication, (iv) a parameter to show order of propagation and (v) the identity of the node. In the quintuple mentioned a combination of the initial three elements are used for computation of reference. Whenever there is a failure of the link, with the help of a these elements a new reference is defined. In the third phase i.e in route erasure phase, this protocol broadcasts a message "clear packet" (CLR) for removing all the routes that are not valid entire the network.

4. ABR

The proposed protocol in [11] does not have loops, free of deadlock and no duplicate packets. The following are the phases of ABR are: (i) finding of Route (ii) If the route is damaged reconstruct the Route and (iii) Removal of Route. In finding of route phase a cycle having Query Broadcast and a REPLY (BQ-REPLY) is used. To transmit packets to target node initially broadcasts a BQ message to find path and searches the nodes which can reach to the target. Based on mobile hosts available on path, the Re-construction of Route phase initiates partial route discovery, removal of invalid routes, update the valid routes and even initiate new routes. When the route discovered is not needed anymore then removal of routes is done as third phase where the initial host broadcasts Delete Route message and all the mobile nodes updates their routing tables.

5. SSR

The working of SSR routing protocol in [16] was mentioned. From his study it can be understood that the "stronger" connectivity is the criteria used in route selection. The author explains that it is a combination of the protocols. The Dynamic Routing Protocol maintains two tables as follows. The Signal Stability Table maintains information of signal's strength of all nearby nodes. It sends tokens at regular intervals from the link layer of each host which is a neighbor node. The strength of the signal based on the SST may be either a strong or weak channel. After all the entries in the table are updated, the message received is passed from DRP to SRP. The role of SRP is to process the messages by sending the message on to the top of the stack if a respective mobile host is the one to receive and if that node is not the node to receive then it searches for target node (T) and then transmits the message. If there is no such a path to reach the target node, then a searching is started to establish the route.

## B. CONCEPTUAL FRAMEWORK

The following is the conceptual framework for this research. A characteristic study of various reactive routing protocols has been conducted. There are various routing strategies adopted by different protocols; of which protocols using multipath routing strategy as shown in the fig. 2.

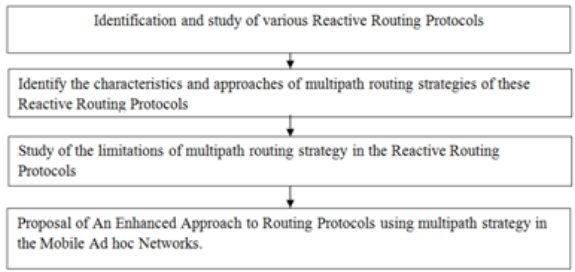

Fig. 2. Conceptual framework

## C. METHODOLOGY

In this paper a comparison of the performance has been conducted between the mobile ad hoc network on demand routing protocols AODV and DSR. We implemented (NS-2) simulator with the following parameters:

TABLE I.  SIMULATION PARAMETERS

| node Number | testing field | Mobile speed | Traffic load | MAC layer | comparison protocols |
|---|---|---|---|---|---|
| 50 | 1500m × 300m | random start to random sink with random speed. | CBR | IEEE 802.11b | DSR & AODV |

## III. SIMULATION RESULT AND DISCUSSION

Packet delivery fraction, Average end-to-end delay of data packets, and Normalized routing load are the performance metrics that have been evaluated. The routing load metric evaluates the efficiency of the routing protocol. The first experiments use varieties number of sources. For the field with 50 nodes the rates of 4 packets per seconds are used. The delivery was the same for booth protocols DSR and AODV when the sources were 10 and 20, but AODV performs when sources were 30 and 40 better than DSR as shown in Fig. 3. AODV doesn't lose more packets like DSR. As shown in Fig. 4. Which it means mobility is higher.

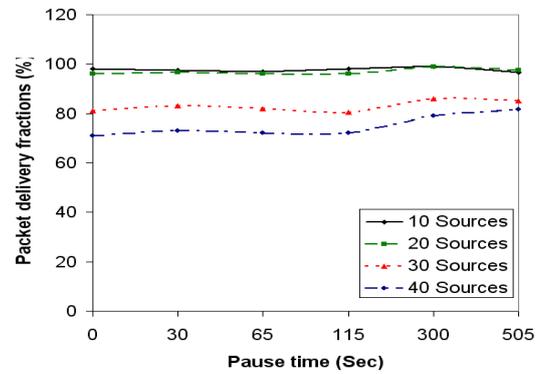

Fig. 3. delivery for the 50 nodes with AODV

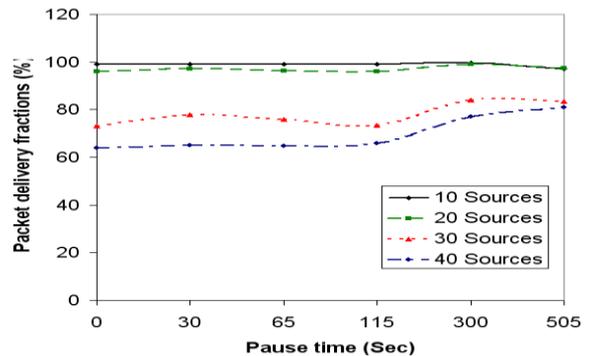

Fig. 4. delivery for the 50 nodes with DSR

DSR has a higher delay than AODV when the sources were between 10 and 20 as shown in Fig. 5. But AODV has less delay with large number of resources as shown in Fig. 6.

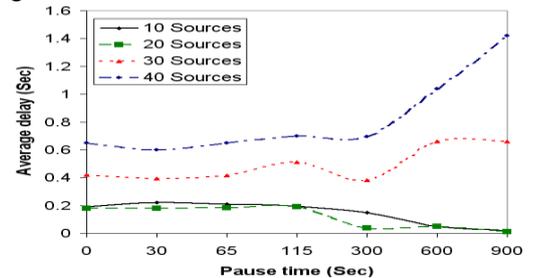

Fig. 5. packet delays for the 50 node in AODV

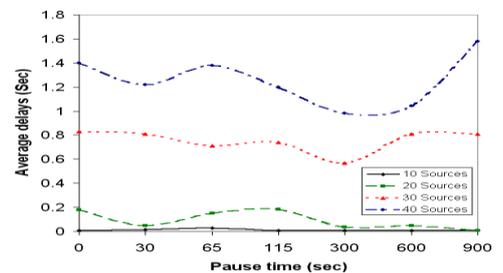

Fig. 6. packet delays for the 50 node in DSR

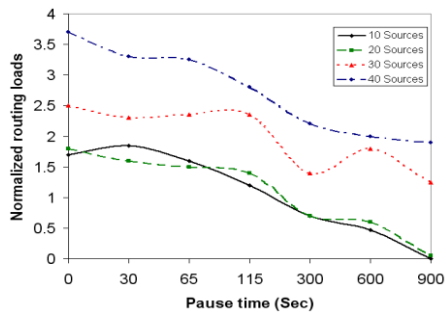

Fig. 7. Normalized loads for the 50 node in AODV

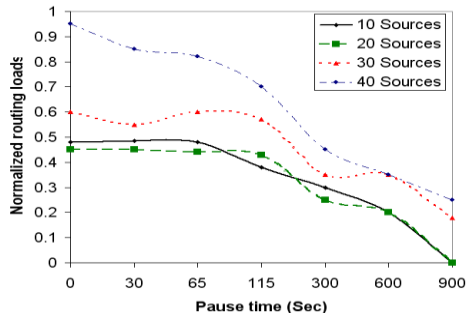

Fig. 8. Normalized loads for the 50 node in DSR

In general AODV demonstrates higher routing load comparing with DSR as shown in Fig. 7. And Fig. 8. , Also booth AODV and DSR were stable when increasing the sources with routing load. For both DSR and AODV were delay when the sources increased by 40 with lower mobility, as shown in Fig. 5. and Fig. 6.

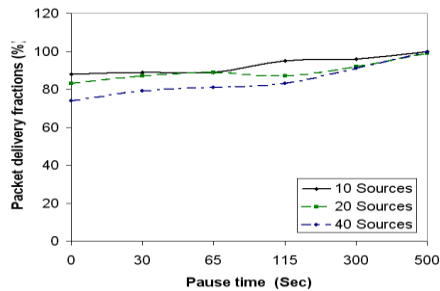

Fig. 9. Packet delivery fractions in AODV.

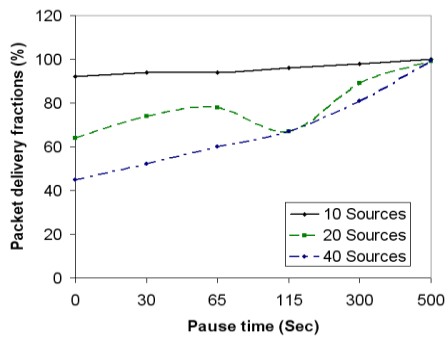

Fig. 10. Packet delivery fractions in DSR.

DSR most of the time has less load than AODV. And performs badly throughput and delay when the field is large and the nodes are many. But it is better than AODV when the field is small and the number of nodes is few.

The routing mechanisms of DSR and AODV are different though they share the on-demand behavior [17]. In particular, source routing is used by DSR, whereas AODV uses table-driven routing and destination sequence numbers. Regarding the timing activities DSR does not rely it whereas AODV relies to a some extent. Studying all these mechanisms and considering the merits of these mechanisms the proposal is to give a new approach to be more effective for these routing protocols. Discovery of route is made in DSR by using route cache table and in AODV by using route table entries in all the intermediate nodes. In both the protocols the discovery of route is made by sending enquiry and reply messages. But the disadvantage is failing to maintain all routes at the same time. Considering this common feature of both the routing protocols a comparison is done and an enhanced approach is proposed to find the route.

The proposal is to add an additional information to the existing Protocol Specific enquiry message proposed by Roy et[18]. In the enquiry message proposed the nodes have the existing information and in addition to that they maintain additional pieces of information like how much amount of current available bandwidth is used, how much amount of battery power remaining etc. This information is required for knowing the neighbor stability information, because the next node's signal strength is also considered as a metric to measure stability of the node. All this information is now used for routing in the network layer. Taking it into consideration an optimal path is selected from all the multiple paths available from a respective node. And then the packet has to be forwarded. During the discovery of route, the routing protocol does not select an optimal route statically but each time some data packet has to be forwarded, it has to dynamically find an optimal route. By dynamically selecting an optimal path here using this new proposal there are two great advantages along with the available traditional multipath routing algorithms. They are having a proper load distribution and using alternate routes. The first advantage of proper load distribution is satisfied because of using the additional information proposed i.e no node is overloaded as the node's stability information is used in the entire network. The second advantage that is met is, for transferring the data the alternate routes can be used resulting in updates thereby preventing the timing out of these routes.

Whenever a particular node has to take a routing decision to forward the packet to the destination the following flowchart is used. In the mentioned flowchart a new search table named Search of Next Node Enquiry Table(SONNET) is proposed to find the best neighbor node. The header of the table has the following contents.

TABLE II. SONNET ATTRIBUTES

| IP Address | MAC Address | Enquiry Message Counter | Time Stamp | Neighbor State | Resource Usage Percentage | Battery Power Left | Signal Strength |
|---|---|---|---|---|---|---|---|
| | | | | | | | |

In this proposal the New Layer is used for finding next best node by periodically sending enquiry messages. Within the radio range whenever a particular node gets an Enquiry message, the fields in the table are updated, which is called Search of Next Node *Enquiry Table* (SONNET). From this new layer, the information like the node's signal strength from the previous node is taken to update the respective field SONNET. Therefore the proposed Search of Next Node Enquiry Table (SONNET) maintains the routing information and helps in dynamically deciding in forwarding the data packet to the best next node.

This study will be a huge contribution in the area of MANETS by using an enhanced approach to multipath routing protocols to reach the destination also which will be helpful for the researchers in designing or developing secure routing protocol in MANETS. The enhanced approach is explained below in the following sub sections

### A. Find multiple paths with loop-free disjoint links

A modification over the Dynamic Source Routing protocol and Adhoc On Demand Distance Vector Routing protocol is proposed. The scheme to discover the routes proposed by Marina et al in [19] is used here. The route discovery process is initiated by using query/reply basis which results in creation of multiple paths. In this scheme loop-free routes are created by making use of *"advertised hop count"* and the multiple paths found ensures link disjointness as proposed in [19] when a single route discovery is initiated.

### B. Format of the Search of Next Node Enquiry Table

The SONNET proposed dynamically finds the next hop for the data packet to be forwarded and the routing entity uses this stability information about the neighboring nodes which is held by SONNET. A New Layer maintains this table, as proposed in [18], by augmenting the protocol specific enquiry messages to carry the information of the node status. In a message which is been sent by the node it contains information of its battery power and the fraction of its available bandwidth currently being used. This message is timely broadcasted to its neighboring nodes considering the range of its antenna. The broadcasting of these messages, not only notifies its presence but also gives more information to its neighbors. The format of the table header is mentioned above. A neighbor is uniquely identified by using the IP address and MAC address which is being recorded. The number of enquiry messages can be known from the EM Counter which is received by the node. The time at which the table was updated last can be known from the Time Stamp. The field Resource Usage Percentage provides the information of how much amount of Node's bandwidth is currently being used. The remaining fields are to give more information of the neighboring node.

### C. Creation and maintenance of the Search of Next Node Enquiry Table

The node when receives an enquiry message from a neighbor, then the data is extracted from it and the SONNET is updated taking into consideration the information related to that respective node. The link layer feedback finds the strength of the signal from a neighbor to know the stability. The Neighbors State is updated following the scheme proposed in [18]. And then periodically the New Layer updates the table SONNET for giving information about the present status of the neighbors. This new approach of DSR or AODV routing protocol helps to take the required information from the SONNET table. From the information and the routing table the nodes having possible multiple paths available can be known so that a decision can be taken to forward the data packet and to which node.

### IV. CONCLUSION AND FUTURE ENHANCEMENTS

The proposed approach intends to mitigate the effects of frequent topological changes in ad hoc networks. A possible optimization to the DSR and AODV routing protocols is proposed to make no node to be overburdened by distributing the load after finding the alternate multipath routes which are discovered in the Route discovery process. A new search table named Search of Next Node Enquiry Table(SONNET) is proposed to find the best neighbor node. This is possible only when an alternate route exists. Using the alternate routes the data is transmitted and the routes are maintained and prevent them from timing out. From the list of neighbors the best next hop to be selected incorporating Qos is proposed in this approach. In future a calculative research can done to select the optimal next hop based on a mathematical analysis from the various values obtained from the SONNET. Also based on the metrics like throughput, communication overhead, Average End-to-End Delay parameters can be used to evaluate these algorithms.